\documentstyle{article}\begin{document}
\title{Schr\"odinger evolution of a scalar field in  Riemannian and
pseudoRiemannian expanding metrics}

\author{ Z. Haba\\
Institute of Theoretical Physics, University of Wroclaw,\\ 50-204
Wroclaw, Plac Maxa Borna 9, Poland,\\
email:zbigniew.haba@uwr.edu.pl}\maketitle
\begin{abstract} We study the
 quantum field theory (QFT) of a scalar field in the  Schr\"odinger picture in the functional formulation.
 We derive a formula
for the evolution kernel in a flat expanding metric. We discuss a
transition between  Riemannian and pseudoRiemannian metrics
$g_{\mu\nu}$ (signature inversion).
 We
express the real time  Schr\"odinger evolution by  the Brownian
motion. We discuss the Feynman integral for a scalar field in a
radiation background. We show that the unitary Schr\"odinger
evolution  for positive time can go over for negative time into a
dissipative evolution  as a consequence of the imaginary value of
$\sqrt{-\det( g_{\mu\nu})}$. The time evolution remains unitary if
$\sqrt{-\det( g_{\mu\nu})}$ in the Hamiltonian is replaced by
$\sqrt{\vert\det( g_{\mu\nu})\vert}$.
\end{abstract}

\section{Introduction}
A quantization of gravity leading to the unified theory of
gravity-matter interaction is still an unsolved problem. The
quantum gravity has its intrinsic problems as  the ultraviolet
divergencies or the notion of time. It may be that time arises
from the Wentzel-Kramers-Brillouin (WKB) semi-classical expansion
of the gravity-matter interaction (see \cite{kiefer} and
references cited there) or from a special form of the
(semi-classical)  energy-momentum \cite{kuchar}. In this letter we
assume that time is well-defined and study the path integral
solution for the scalar field wave function evolution
$\psi_{t}(\Phi)$ in an external metric $g_{\alpha\beta}$
\begin{equation} \psi_{t}(\Phi) =\int {\cal
D}\Phi(.)\exp\Big(\frac{i}{\hbar}\int dx\sqrt{- g}{\cal
L}(g_{\mu\nu},\Phi)\Big)\psi_{0}(\Phi_{t}(\Phi)),
\end{equation}
where  $g=\det(g_{\mu\nu})$, ${\cal L}$ is the Lagrangian,
$\Phi_{t}(\Phi)$ is the path starting from $\Phi$ at $t=0$ and
$\psi_{0}(\Phi)$ is the initial state. We consider the path
integral (1) for both Lorentzian and Euclidean metrics. We
encountered the path integral (1) in the real time and with an
inverted signature  in the upside-down oscillator (${\cal L}\simeq
(\frac{dq}{dt})^{2}+\nu^{2}q^{2}, \nu^{2}>0$) which behaves like a
free field theory with Euclidean signature
\cite{hababook}\cite{habarxiv} \cite{guth}\cite{barton}.

 For a free field theory (quadratic
Lagrangian) the evolution (1) has been studied by many authors
(see e.g. \cite{fulling} and references cited there). In this
paper we obtain a formula for the time evolution (1) in the case
of a homogeneous time-dependent metric (see some earlier results
in \cite{prokop1}\cite{prokop2}\cite{parker}). The solution of the
Schr\"odinger equation is expressed by a stochastic process (a
Gaussian field). Using the time evolution for the free field we
can derive the time evolution for an interacting field by means of
the Feynman-Kac formula. The time evolution can be calculated in
perturbation theory using the free field evolution kernel.
 We assume the initial state $\psi_{0}=
\exp(\frac{i}{\hbar}S_{0})\chi$ in the WKB form. We show that if
the initial state  is Gaussian ($\chi=1$) then $\psi_{t}$ is also
Gaussian. We expand the evolution of $\psi_{0}=
\exp(\frac{i}{\hbar}S_{0})\chi$ around the Gaussian state.We
expect the caustic problems in the evolution kernel as they appear
already in the Mehler formula for an oscillator. The problem can
be approached  for a small time by an ultraviolet regularization (
or a finite number of modes) and a subsequent continuation in time
\cite{maslov}.

 In this letter after a description in secs.2-3 of the Schr\"odinger evolution
 in the functional (stochastic) formulation (based on \cite{habaepj}\cite{hababook})
  we concentrate on the model of a quantum scalar field in a background of radiation.
  The spatial metric
$g_{jk}= \delta_{jk}a^{2}\simeq \delta_{jk}t$ becomes degenerate
at $t=0$ and changes signature when $t<0$.
 The solution of
Einstein equations for radiation with $a^{2}=t<0$ is usually
rejected as unphysical \cite{landau} (sec.112).  However, it
should be taken into account in quantum gravity when the
functional integral (1) is over all metrics. The inverted
signature can appear in functional integration as a saddle point
in eq.(1). Then, a differentiation of eq.(1) over time will lead
to the Schr\"odinger equation with an inverted  signature. We
discuss in detail the Schr\"odinger evolution (1) for various WKB
initial states. We show that for some WKB states the unitary
Schr\"odinger evolution for positive time can go over the
singularity at time zero in a continuous way into a dissipative
evolution described by diffusive paths. We show that the time
evolution can remain unitary if $\sqrt{-\det( g_{\mu\nu})}$ in
eq.(1) is replaced by $\sqrt{\vert\det( g_{\mu\nu})\vert}$. We
compare this behavior with the one for $a^{2}\simeq \vert t\vert $
derived in the string inspired models \cite{string}.

The model studied in this letter delivers an example of the
Gibbons-Hartle-Hawking suggestion \cite{hartle}\cite{gibbons} of a
continuation of the time evolution by means of the path integral
to the region of the Euclidean signature. In our model the
signature is changing for a negative time (see  similar models in
\cite{ellis}\cite{visser1}). In a forthcoming paper \cite{habanew}
we discuss some other models with an interaction and a signature
inversion resulting from a continuation in time.
        \section{WKB solution of the Schr\"odinger equation}
We restrict ourselves to the metric
\begin{equation}
ds^{2}=g_{\mu\nu}dx^{\mu}dx^{\nu}\equiv
g_{00}dt^{2}-g_{jk}dx^{j}dx^{k} =dt^{2}-a^{2}d{\bf x}^{2}.
\end{equation}
  The free scalar field on the background metric (2) satisfies the equation \begin{equation}
  \partial_{t}^{2}\Phi-a^{-2}\triangle\Phi+3H\partial_{t}\Phi+m^{2}\Phi=0,
  \end{equation}where
  $H=a^{-1}\partial_{t}a$.

The Lagrangian for the system (3) is
\begin{equation}
{\cal
L}=\frac{1}{2}\sqrt{-g}\Big(g^{\mu\nu}\partial_{\mu}\Phi\partial_{\nu}\Phi-m^{2}\Phi^{2}\Big),
\end{equation}
the canonical momentum \begin{equation}
\Pi=\sqrt{-g}\partial_{t}\Phi
\end{equation}
 and the Hamiltonian
\begin{equation}\begin{array}{l}
{\cal H}(t)=\int d{\bf x}\sqrt{-g}{\cal H}(t,{\bf
x})\cr=\frac{1}{2}\int d{\bf x}\Big(
a^{-3}\Pi^{2}+a(\nabla\Phi)^{2}+m^{2}a^{3}\Phi^{2}\Big).\end{array}
\end{equation}
Using the representation $\Pi({\bf
x})=-i\hbar\frac{\delta}{\delta\Phi({\bf x})}$ we obtain
\begin{equation} {\cal H}(t)=\frac{1}{2}\int d{\bf x}
\Big(-\hbar^{2}a^{-3}\frac{\delta^{2}}{\delta\Phi({\bf
x})^{2}}+a(\nabla\Phi)^{2}+m^{2}a^{3}\Phi^{2}\Big).
\end{equation}
We consider a time-dependent Gaussian solution of the
Schr\"odinger equation
\begin{equation}i\hbar\partial_{t}\psi_{t}={\cal H}(t)\psi_{t}.
\end{equation}

   The expression for the Feynman path integral solution of the
   Schr\"odinger equation (8) with the initial condition
     \begin{equation}
     \psi^{g}_{0}=\exp(\frac{i}{\hbar}S_{0})
   \end{equation}
   in a fixed background metric has the form
   \begin{equation} \begin{array}{l}
   \psi_{t}^{g}(\Phi)=\int d\Phi(.)\exp\Big(\frac{i}{2\hbar}\int_{0}^{t} ds d{\bf
   x}\Big(\sqrt{-
   g}g^{\mu\nu}\partial_{\mu}\Phi\partial_{\nu}\Phi+m^{2}\sqrt{- g}\Phi^{2}\Big)\cr
   \times \exp(\frac{i}{\hbar}S_{0}(\Phi_{t}(\Phi)).
   \end{array}\end{equation} In this letter we restrict ourselves to a real
   function
    $S_{0}$  quadratic  in $\Phi$.
   We expand the Feynman integral (10) around the stationary point
   $\phi_{s}^{c}(\Phi)$. The solution of the Cauchy problem
   $\phi_{s}^{c}(\Phi)$ with the initial field value $\Phi$
and the final boundary condition on the derivative
\begin{equation}
\frac{d\phi_{t}^{c}}{dt}=-\frac{\delta
S_{0}(\phi)}{\delta\phi}(\phi_{t}^{c})
\end{equation}
   is linear in $\Phi$. We write ($ s\leq t$) \begin{equation}
   \Phi_{s}=\phi_{s}^{c}(\Phi)+\sqrt{\hbar}\phi_{q}.
   \end{equation}
   Then,
   \begin{equation}
   \psi_{t}^{g}(\Phi)=A_{t}\exp(\frac{i}{\hbar}S_{t}(\phi_{c}(\Phi))
   \equiv A_{t}\exp(\frac{i}{2\hbar}(\Phi,\Gamma_{t}\Phi)),\end{equation}
   where $(,)$ denotes  the scalar product in $L^{2}(d{\bf x})$.

\section{Expansion around the WKB solution}
We write a general solution of the Schr\"odinger equation (8)in
the form
\begin{equation}
\psi_{t}=\psi_{t}^{g}\chi_{t}.
\end{equation}
 Then, $\chi$ solves the
equation
\begin{equation} i\hbar\partial_{t}\chi=\frac{1}{2}\int d{\bf
x}a^{-3}\Big( \Pi^{2}+(\Pi\ln\psi_{t}^{g})\Pi\Big)\chi.
\end{equation}
For  the solution (13) eq.(15) reads
\begin{displaymath}\partial_{t}\chi_{t}= \int d{\bf x}
\Big(i\hbar a^{-3}\frac{1}{2}\frac{\delta^{2}}{\delta\Phi({\bf
x})^{2}}-a^{-3}\Gamma\Phi({\bf x})\frac{\delta}{\delta\Phi({\bf
x})}\Big)\chi.
\end{displaymath}

Eq.(15) is a diffusion equation in an infinite number of
dimensions \cite{daletski} with an imaginary diffusion constant .
If $\chi$ is a holomorphic function then the  solution of eq.(15)
can be expressed as an expectation value
\cite{freidlin}\cite{habaepj}

 \begin{equation}
\chi_{t}(\Phi)=E\Big[\chi\Big(\Phi_{t}(\Phi)\Big)\Big].
\end{equation}
In eq.(16) $\Phi_{s}(\Phi)$  for $0\leq s\leq t$ is the solution
of
 the stochastic Langevin equation \cite{freidlin}
(  this equation holds true for an arbitrary solution
$\psi_{t}^{g}$ of eq.(8))
 \begin{equation}
 d\Phi_{s}({\bf x})=i\hbar a(t-s)^{-3}\frac{\delta}{\delta \Phi_{s}({\bf
 x})}\ln\psi_{t-s}^{g}ds+\sqrt{i\hbar}a(t-s)^{-\frac{3}{2}}dW_{s}({\bf x})
 \end{equation}
 with the initial condition $\Phi$. In eq.(17)
 $\sqrt{i}=\exp(i\frac{\pi}{4})=\frac{1}{\sqrt{2}}(1+i)$.
The Brownian motion is defined (for non-negative time) as the
Gaussian process with mean zero and the correlation function
\begin{equation} E[W_{s}({\bf x})W_{s^{\prime}}({\bf
    y})]=min(s,s^{\prime})\delta({\bf x}-{\bf y}).\end{equation}
We denote the Fourier transform of $W$ by the same letter. It has
the covariance
\begin{displaymath} E[W_{s}({\bf k})W_{s^{\prime}}({\bf
    k}^{\prime})]=min(s,s^{\prime})\delta({\bf k}+{\bf k}^{\prime}).\end{displaymath}
 Eq.(17) for the solution (13) takes the form
 \begin{equation}d\Phi_{s}=-a(t-s)^{-3}\Gamma(t-s)\Phi_{s}
 ds+a(t-s)^{-\frac{3}{2}}\sqrt{i\hbar}dW_{s}.
 \end{equation} In eq.(19) $\Gamma$ is an operator acting upon $\Phi$. In Fourier transforms in the next sections
 $\Gamma$ will be a multiplication operator.
The solution of the stochastic equation (17) determines the free
 field correlation functions in the state $\psi_{0}^{g}\chi$
($\hat{\Phi}$ denotes the quantum field)
\begin{equation}\begin{array}{l}(\psi_{0}^{g}\chi,F_{1}(\hat{\Phi}_{t})F_{2}(\hat{\Phi})\psi_{0}^{g}\chi)
\cr=\int d\Phi \vert\psi_{t}^{g}(\Phi)\vert^{2}F_{1}(\Phi)
E\Big[\chi\Big(\Phi_{t}(\Phi)\Big)\Big]^{*}
E\Big[F_{2}\Big(\Phi_{t}(\Phi)\Big)\chi\Big(\Phi_{t}(\Phi)\Big)\Big].
\end{array}\end{equation}

\section{Free field in an expanding
universe} We can determine the operator $\Gamma_{t}$ in eq.(13)
inserting $\psi_{t}^{g}$ into the Schr\"odinger equation (8).
$\psi_{t}^{g}$ solves the Schr\"odinger equation (8) if $\Gamma$
satisfies
 \begin{displaymath}
 \partial_{t}\Gamma+a^{-3}\Gamma^{2}+(-a\triangle +m^{2}a^{3})=0.
 \end{displaymath}
We can relate this non-linear Riccati equation to a linear second
order equation if we introduce the operator $u$
\begin{displaymath}
u=\exp(\int^{t}ds a(s)^{-3}\Gamma_{s}).
\end{displaymath}
Then,
\begin{displaymath}
\frac{d^{2}u}{dt^{2}}+3H\frac{du}{dt}+(-a^{-2}\triangle+m^{2})u=0.
\end{displaymath}
We obtain $\Gamma$ as
\begin{displaymath}
\Gamma=u^{-1}\frac{du}{dt}a^{3}.
\end{displaymath}
Equations for $\Gamma$ and $u$ can be derived for a general
metric. However, explicit  solutions are available only if the
metric has a large group of symmetries (de Sitter group will be
discussed in \cite{habanew}). In our homogeneous case the group of
translations allows to apply the Fourier transform
 \begin{displaymath}
\Gamma({\bf x}-{\bf y})=(2\pi)^{-3}\int d{\bf k}\Gamma({\bf
k})\exp(i{\bf k}({\bf x}-{\bf y})).
\end{displaymath}
 If $a^{2}>0$ then we consider solutions satisfying
the reality condition in the configuration space (then in the
Fourier space $\Gamma^{*}({\bf k})=\Gamma(-{\bf k})) $ as well as
in the Fourier space (we choose $\Gamma({\bf k})=\Gamma(-{\bf k})
=\Gamma(k)$, where $k=\vert{\bf k}\vert$). In Fourier space the
equation for $\Gamma$ reads \begin{equation}
\partial_{t}\Gamma+a^{-3}\Gamma^{2}+a{\bf k}^{2}+m^{2}a^{3}=0.
\end{equation}
Eq.(21) can be related to the second order differential equation
(this is the wave equation (3) in the  momentum space)

\begin{displaymath}
\frac{d^{2}u}{dt^{2}}+3H\frac{du}{dt}+(a^{-2}{\bf
k}^{2}+m^{2})u=0.
\end{displaymath}
The stochastic equation (19) has the solution (with the initial
condition  $\Phi$ at $s=0$)
 \begin{equation}
 \Phi_{s}(\Phi)=u_{t-s}u_{t}^{-1}\Phi+\sqrt{i\hbar}u_{t-s}\int_{0}^{s}u_{t-\tau}^{-1}a_{t-\tau}^{-\frac{3}{2}}dW_{\tau}.
 \end{equation}
The Schr\"odinger equation (8) is considered for positive as well
as for negative time whereas the solution (16) is defined only for
a positive time. We can extend this definition to the negative
time using the time reflection symmetry of eq.(8)
\begin{equation}i\hbar\partial_{-t}\psi_{t}^{*}={\cal H}^{*}(t)\psi_{t}^{*}=
\tilde{{\cal H}}(-t)\psi_{t}^{*},
\end{equation}where $\tilde{{\cal H}}(-t)={\cal H}^{*}(t)$.
The operator $\tilde{{\cal H}}(-t)$ can be different then ${\cal
H}(t)$ as in the model of sec.5 when $\sqrt{-g}$ becomes imaginary
for negative time. Nevertheless, the Feynman formula (1) gives the
solution of eq.(23) as will be discussed in the model of sec.5.
 In
the simplest case when ${\cal H}(t)={\cal H}(-t)$ and $a(t)=
a(-t)$ (as in the CPT model
 \cite{cpt} with $a^{2}(t)=\vert t \vert $ ) the reflection symmetry  means  that in order to define the
evolution for $t<0$ we apply the stochastic equation (19) for
$-t\geq -s\geq 0$
\begin{displaymath}
d\Phi_{s}=\Gamma(s-t)a^{-3}(t-s)\Phi_{s}d(-s)+\sqrt{-i\hbar}a^{-\frac{3}{2}}(t-s)dW_{-s}.
\end{displaymath}
In this equation
  $\Gamma\rightarrow -\Gamma$  because of the complex
conjugation of $\psi_{t}^{g}$ with a real $\Gamma$ in the
representation (14) of $\psi_{t}$  and (for the same reason) we
have a complex conjugation of $\sqrt{i\hbar}$. So, the solution of
the Schr\"odinger equation for the negative time $t<0$ is
$\chi_{t}(\Phi)=E[\chi(\Phi_{t}(\Phi))]$, where
\begin{equation} \Phi_{s}(\Phi)=u^{(-)}_{t-s}(u^{(-)}_{t})^{-1}\Phi
-\sqrt{-i\hbar}
u^{(-)}_{t-s}\int_{0}^{-s}(u^{(-)}_{t+\tau})^{-1}a_{t+\tau}^{-\frac{3}{2}}dW_{\tau},
\end{equation}
where $u_{t}^{(-)}$ is the solution of the wave equation (3) for
the negative time. Eq.(19) is related to the semi-classical
representation (13). By differentiation of eq.(19) and using
eq.(21) we obtain a random wave equation
\begin{displaymath}\begin{array}{l}
(\partial_{s}^{2}-a^{-2}\triangle
+m^{2})\Phi_{s}+3a^{-1}(t-s)\partial_{t}a(t-s)\partial_{s}\Phi_{s}
\cr=
(\frac{9}{2}a^{-1}(t-s)\partial_{t}a(t-s)\cr-a^{-3}(t-s)\Gamma(t-s))\sqrt{i\hbar}a^{-\frac{3}{2}}(t-s)\partial_{s}W
+\sqrt{i\hbar}a^{-\frac{3}{2}}(t-s)\partial^{2}_{s}W.
\end{array}\end{displaymath}

 The
evolution kernel is defined by
\begin{equation}
(U_{t}\psi)(\Phi)=\int d\Phi^{\prime}
\tilde{K}_{t}(\Phi,\Phi^{\prime})\psi(\Phi^{\prime}).
\end{equation}
We write $\psi$ in the form (14) and rewrite eq.(25) as
\begin{equation}
(U_{t}\psi_{0}^{g}\chi)(\Phi)=\psi_{t}^{g}\int d\Phi^{\prime}
K_{t}(\Phi,\Phi^{\prime})\chi(\Phi^{\prime}).
\end{equation}
We can derive the evolution kernel explicitly  assuming a Fourier
representation of $\chi$
\begin{displaymath}
\chi(\Phi)=\int d\chi \tilde{\chi}(\Lambda)\exp(i(\Lambda,\Phi)).
\end{displaymath}We obtain ( using eq.(22))
\begin{equation}\begin{array}{l}\chi_{t}(\Phi)=\int d\chi \tilde{\chi}(\Lambda)
E\Big[\exp(i(\Lambda,\Phi_{t}(\Phi))\Big] \cr=\int d\chi
\tilde{\chi}(\Lambda)\exp\Big(i(\Lambda,u_{0}u_{t}^{-1}\Phi)-\frac{1}{2}
(\Lambda,G_{t}\Lambda)\Big),\end{array}\end{equation}
 where
\begin{equation}\begin{array}{l}
            G_{t}({\bf k},{\bf k}^{\prime})=E[(\Phi_{t}({\bf k})-E[\Phi_{t}]({\bf k}))(\Phi_{t}({\bf k}^{\prime})
            -E[\Phi_{t}]({\bf k}^{\prime}))]\cr
=i\delta({\bf k}+{\bf k}^{\prime})\hbar
u_{0}^{2}\int_{0}^{t}u_{t-\tau}^{-2}a_{t-\tau}^{-3}d\tau
\end{array}\end{equation} (eq.(28) results from the white noise
correlations of $\frac{dW}{d\tau}$).

Then, it follows
\begin{equation}\begin{array}{l}
K_{t}(\Phi,\Phi^{\prime}) =\int d\Lambda
\exp\Big(i(\Lambda,u_{0}u_{t}^{-1}\Phi-\Phi^{\prime}))-\frac{1}{2}
(\Lambda,G_{t}\Lambda)\Big).\end{array}\end{equation} The Gaussian
integral (29) gives an explicit formula for the evolution kernel
in terms of the implicit function $G_{t}$  ( this is another
solution of the problem posed in \cite{prokop1}\cite{prokop2}).
 Eq.(29) can also be used to check by means of
a direct calculation that the formula (16) gives the solution of
eq.(15). We calculate $G_{t}$ in the next section.

    \section{ The signature inversion in the evolution $ a^{2}(t)=c_{0}^{-1}t$}
     We consider a massless  ($m=0$ in eq.(3)) scalar field and the spatial metric
     $g_{lk}=\delta_{lk}c_{0}^{-1}t$ ($c_{0}>0$). This metric with  $t\in R$ is the solution of the Friedmann equations for
     radiation with the energy density  $\rho=\rho_{0}a^{-4}$ and
     the pressure $p=\frac{1}{3}\rho$. At $t=0$ both sides of
     Friedmann equations are infinite as at the Big Bang the
     curvature tensor and the energy-momentum of radiation are
     infinite. We assume that the metric   $g_{lk}=\delta_{lk}c_{0}^{-1}t$ can apper as a saddle point in the functional integral for quantum gravity.
     It would need a  detailed investigation in concrete models of radiation to determine whether
     the classical action is finite for the solution $ a^{2}(t)=c_{0}^{-1}t$.
     When we differentiate eq.(1) over $t$ then we obtain the Schr\"odinger equation (8) for
     $t>0$ as well as for $t<0$. The initial value problem for the
     Schr\"odinger equation is usually formulated at $t=0$. In
     order to separate the initial value for the wave function
     from the initial value for classical Einstein equations we shift the
     time choosing $ a^{2}(t)=c_{0}^{-1}(t+\gamma)$.
     We insert $\gamma\geq 0$ for $t\geq 0$ and $\gamma\leq 0$ for $t\leq
     0$ so that when $\gamma\neq 0$ the metric is not degenerate (  $\gamma$ is treated as a regularization).

     For $t+\gamma<0$ when $a^{2}<0$ then we choose $a=ic_{0}^{-\frac{1}{2}}\sqrt{\vert t+\gamma\vert}$. The
     Schr\"odinger equation (8) reads
\begin{displaymath}\begin{array}{l}
\hbar\partial_{t}\psi_{t}= \frac{1}{2}\int d{\bf x}
\Big(-\hbar^{2}c_{0}^{\frac{3}{2}}\vert
t+\gamma\vert^{-\frac{3}{2}}\frac{\delta^{2}}{\delta\Phi({\bf
x})^{2}}+c_{0}^{-\frac{1}{2}}\vert
t+\gamma\vert^{\frac{1}{2}}(\nabla\Phi)^{2}-m^{2}c_{0}^{-\frac{3}{2}}\vert
t+\gamma\vert^{\frac{3}{2}}\Phi^{2}\Big)\psi_{t}.\end{array}
\end{displaymath}
The equation for $\chi$ is
\begin{equation}\begin{array}{l}\partial_{t}\chi_{t}= \int d{\bf x}
\Big(-\frac{\hbar}{2}c_{0}^{\frac{3}{2}}\vert
t+\gamma\vert^{-\frac{3}{2}}\frac{\delta^{2}}{\delta\Phi({\bf
        x})^{2}}-(u^{(-)})^{-1}\partial_{t}u^{(-)} \Phi({\bf
    x})\frac{\delta}{\delta\Phi({\bf x})}\Big),\end{array}
\end{equation}where $u^{(-)}$ is the solution of the wave equation for  $t+\gamma<0$.
Let us note that the drift term  in eq.(30) depends only on $u$
and does not depend directly on the signature of $a^{2}$ . As a
consequence of the purely imaginary $a$ for the inverted signature
we obtain a diffusion equation with a real diffusion constant.

It is instructive to compare the quantum scalar field in the
metric $a^{2}\simeq t$ with the one for $a^{2}\simeq \vert t\vert
$ (this non-analytic metric cannot be a solution of Einstein
equations for $t\in R$ but does appear in string-inspired models
\cite{string}). In the latter case the Schr\"odinger equation for
$\chi$ reads
\begin{displaymath}\begin{array}{l}\partial_{t}\chi_{t}= \int d{\bf x}
\Big(i\hbar\frac{1}{2}c_{0}^{\frac{3}{2}}\vert
t+\gamma\vert^{-\frac{3}{2}}\frac{\delta^{2}}{\delta\Phi({\bf
        x})^{2}}-\tilde{u}^{-1}\partial_{t}\tilde{u} \Phi({\bf
    x})\frac{\delta}{\delta\Phi({\bf x})}\Big),\end{array}
\end{displaymath}
where $\tilde{u}$ is the solution of the wave equation
\begin{displaymath}
\frac{d^{2}\tilde{u}}{dt^{2}}+\frac{3}{2}(t+\gamma)^{-1}\frac{d\tilde{u}}{dt}
+\vert  t+\gamma\vert^{-1}c_{0}k^{2}\tilde{u}=0.
\end{displaymath}
 This equation is invariant under the time reflection so
$ \tilde{u}(-t-\gamma)=\tilde{u}(t+\gamma)$. Hence, also
 $\Gamma$  is invariant under time reflection. As a consequence,
 the field $\Phi_{s}$ for a negative time will be just a
 reflection of the one for positive time.
  The solution $\tilde{u}$ of the wave equation can be obtained if we change the
 cosmic time $t$ in eq.(2) into the conformal time  $T$ as $T=\int dt a^{-1}=2c_{0}^{\frac{1}{2}}\sqrt{t}$ for $ t>0$ and
 $T=-2c_{0}^{\frac{1}{2}}\sqrt{-t}$ for $t<0$ ($a^{2}(T)=\frac{1}{4c_{0}}T^{2}$ as in \cite{cpt}).
  Then,
 $v=T\tilde{u}$ satisfies the oscillator equation. Hence, the
 solution of the wave equation with the metric $a^{2}\simeq \vert t\vert $ is a superposition of plane
 waves in the conformal time (with a time-dependent decaying amplitude)
 \begin{displaymath}
\tilde{u}=A_{1}T^{-1}\exp(ikT)+A_{2}T^{-1}\exp(-ikT).
 \end{displaymath}
 This will not be so for the metric $a^{2}\simeq t$ for negative time because if $a^{2}<0$ then instead
 of the hyperbolic equation (3) we obtain an elliptic equation.
    The field $\Phi_{s}$ for a negative $s$ will be a diffusion
    process instead of an oscillatory one for $s>0$.
     The Gaussian solution (13) is determined for positive as well for negative time by  a solution of the equation
     \begin{equation}
     \frac{d^{2}u}{dt^{2}}+\frac{3}{2}(t+\gamma)^{-1}\frac{du}{dt}
     +(t+\gamma)^{-1}c_{0}k^{2}u=0
     \end{equation}
      true for   $t+\gamma>0$ as well as $t+\gamma<0$.
      Note that eq.(31) at  $t+\gamma<0$ does not describe a wave
      propagation but rather a damped oscillator with the imaginary frequency $\omega^{2}<0$.
     We express the solution of eq.(31) in terms of real functions
       \begin{displaymath}\begin{array}{l}
       u_{t}=C_{1}(t+\gamma)^{-\frac{1}{2}}\cos(2k\sqrt{c_{0}}\sqrt{t+\gamma})\cr+C_{2}(t+\gamma)^{-\frac{1}{2}}\sin(2k\sqrt{c_{0}}\sqrt{t+\gamma})
       \end{array}\end{displaymath}
       if $t+\gamma>0$ and
          \begin{displaymath}\begin{array}{l}
          u^{(-)}_{t}=C_{1}(-t-\gamma)^{-\frac{1}{2}}\cosh(2k\sqrt{c_{0}}\sqrt{-t-\gamma})+C_{2}(-t-\gamma)^{-\frac{1}{2}}\sinh(2k\sqrt{c_{0}}\sqrt{-t-\gamma})
         \end{array} \end{displaymath}
if $t+\gamma<0$.

The solutions of the wave equation (31) determine also the
solution of the Schr\"odinger equation with $\sqrt{-g}$ replaced
by $\sqrt{\vert g\vert}$ in the Hamiltonian (6). Then, the
equation for $\chi$ at negative time reads
\begin{equation}\begin{array}{l}\partial_{t}\chi_{t}= \int d{\bf x}
\Big(i\hbar\frac{1}{2}c_{0}^{\frac{3}{2}}\vert
t+\gamma\vert^{-\frac{3}{2}}\frac{\delta^{2}}{\delta\Phi({\bf
        x})^{2}}-(u^{(-)})^{-1}\partial_{t}u^{(-)} \Phi({\bf
    x})\frac{\delta}{\delta\Phi({\bf x})}\Big),\end{array}
\end{equation}
where $u^{(-)}$ is the solution of eq.(31) for negative time.

We are interested in the behavior of the quantum scalar field
evolution in the limit $\gamma=0$.
 The limit $\gamma\rightarrow 0$ of $u_{t}$
  exists for all $\vert t\vert\geq 0$ (as required in classical field theories with a signature change
 \cite{ellis}\cite{visser1}\cite{visser2} \cite{visser}) only if $C_{1}=0$. Then, the limit $t+\gamma \rightarrow 0$ for
 positive time as well as  for the negative time is the same
 $u_{0}=2C_{2}\sqrt{c_{0}}k$.
 The limit  $t+\gamma\rightarrow 0$ of $
 \partial_{t}u_{t}$ also exists from both sides
 \begin{displaymath}
(\partial_{t}u_{t})_{\vert
t=0}=-\frac{4}{3}C_{2}\sqrt{c_{0}}c_{0}k^{3}.
\end{displaymath}

        The solution of the stochastic equation (19) for $t\geq
        s\geq 0$ is
        \begin{equation}\begin{array}{l}
        \Phi_{s}(\Phi)=u_{t-s}u_{t}^{-1}\Phi+\sqrt{i\hbar}u_{t-s}c_{0}^{\frac{3}{4}}\int_{0}^{s}u_{t-\tau}^{-1}
        \vert t+\gamma-\tau\vert^{-\frac{3}{4}}dW_{\tau}.
        \end{array}\end{equation}
        The quantum field theory is defined by  the correlation  function (for a
        positive time)
       \begin{equation}\begin{array}{l}
            G_{ss^{\prime}}=E[(\Phi_{s}({\bf k})-E[\Phi_{s}]({\bf k}))(\Phi_{s^{\prime}}({\bf k}^{\prime})
            -E[\Phi_{s^{\prime}}]({\bf k}^{\prime}))]\cr
            =\delta({\bf k}+{\bf k}^{\prime})i\hbar c_{0}^{\frac{3}{2}}u_{t-s}u_{t-s^{\prime}}\int_{0}^{m(s,s^{\prime})}d
            \tau
              u_{t-\tau}^{-2}   \vert
              t+\gamma-\tau\vert^{-\frac{3}{2}}\cr\equiv \delta({\bf k}+{\bf k}^{\prime})   G_{ss^{\prime}}(k).\end{array}
             \end{equation}We denote $m(s,s^{\prime})\equiv min(s,s^{\prime})$ if $t\geq s\geq 0$
             and $t\geq s^{\prime}\geq 0$.
             The expectation value (34) comes from the
             correlation $E[\frac{dW}{d\tau}({\bf k})\frac{dW}{d\tau^{\prime}}({\bf
             k^{\prime}})]
             =\delta(\tau-\tau^{\prime})\delta({\bf k}+{\bf
             k}^{\prime})$. The $\delta({\bf k}+{\bf
             k}^{\prime})$ term will be omitted  in the formulas below.

For a negative time driven by the diffusion equation (30) there is
no $\sqrt{i}$ in eq.(33) which is cancelled by the   $\sqrt{i}$
factor in $a^{-\frac{3}{2}}$. Hence, the counterpart of eq.(33)
for the negative time reads
\begin{equation}\begin{array}{l}
        \Phi_{s}(\Phi)=u^{(-)}_{t-s}(u^{(-)}_{t})^{-1}\Phi+\sqrt{\hbar}u^{(-)}_{t-s}c_{0}^{\frac{3}{4}}
        \int_{0}^{-s}(u^{(-)}_{t+\tau})^{-1}
        \vert t+\gamma+\tau\vert^{-\frac{3}{4}}dW_{\tau}.
        \end{array}\end{equation}
If the time evolution  for negative time is determined by eq.(32)
(resulting from the replacement of $\sqrt{-g}$ by $\sqrt{\vert
g\vert}$ in the Hamiltonian (6)) then in the solution $\Phi_{s}$
in eq.(35) we replace $ \sqrt{\hbar}$ by $\sqrt{-i\hbar}$ as in
eq.(24).

We calculate the correlation function of the fields (35) for a
negative time
             \begin{equation}\begin{array}{l}
            G_{ss^{\prime}}=E[(\Phi_{s}({\bf k})-E[\Phi_{s}]({\bf k}))(\Phi_{s^{\prime}}({\bf k}^{\prime})
            -E[\Phi_{s^{\prime}}]({\bf k}^{\prime}))]\cr
            =\delta({\bf k}+{\bf k}^{\prime})\hbar c_{0}^{\frac{3}{2}}u^{(-)}_{t-s}u^{(-)}_{t-s^{\prime}}
            \int_{0}^{m(s,s^{\prime})}d
            \tau
              (u^{(-)}_{t-\tau})^{-2}   \vert
              t+\gamma+\tau\vert^{-\frac{3}{2}}\equiv \delta({\bf k}+{\bf k}^{\prime})   G_{ss^{\prime}}(k).\end{array}
             \end{equation}where for the negative time $m(s,s^{\prime})\equiv min(-s,-s^{\prime})$ if $-t\geq -s\geq 0$ and $-t\geq -s^{\prime}\geq 0$.

It follows that for a negative time $\Phi_{s}$ becomes a real
diffusion process and $K_{t}$ in eq.(29) is a real transition
function. The fact that $\Phi_{s}$ for a negative time is a  real
diffusion process follows already from eq.(30) which is an
equation for a diffusion  with a real diffusion constant and a
real drift.

         We can obtain an explicit formula (expressed by elementary functions) for the
         correlation functions (34) and (36)
                  if   $C_{2}=0$, $C_{1}=0$ or $C_{1}=\pm C_{2}$. We are interested in the calculation of
         $G_{ss^{\prime}}$ (34)-(36) for all these cases
         although  some of the
         solutions $u_{t}$ may have no limit when $t+\gamma$ is equal to zero .

         If $C_{2}=0$ then we
         have for $t+\gamma >0$
         \begin{equation}
         u_{t}=(t+\gamma)^{-\frac{1}{2}}\cos(2\sqrt{c_{0}}k\sqrt{t+\gamma}).
         \end{equation}
         If $C_{1}=0$
         \begin{equation}
         u_{t}=(t+\gamma)^{-\frac{1}{2}}\sin(2\sqrt{c_{0}}k\sqrt{t+\gamma}).
         \end{equation}
          If  $ -t-\gamma>0$ then
        the corresponding formulas read
            \begin{equation}    u^{(-)}_{t}=(-t-\gamma)^{-\frac{1}{2}}\cosh(2\sqrt{c_{0}}k\sqrt{-t-\gamma})
            \end{equation}
        and
              \begin{equation}
              u^{(-)}_{t}=(-t-\gamma)^{-\frac{1}{2}}\sinh(2\sqrt{c_{0}}k\sqrt{-t-\gamma}).
            \end{equation}

          When $C_{1}=\pm C_{2}$
             \begin{equation}
               u_{t}^{(-)}=(-t-\gamma)^{-\frac{1}{2}}\exp(\pm
               2\sqrt{c_{0}}k\sqrt{-t-\gamma}).
              \end{equation}
            For the solution (37) we obtain
              \begin{equation}\begin{array}{l}
              G_{ss^{\prime}}=i\hbar u_{t-s}u_{t-s^{\prime}}
              c_{0}^{\frac{3}{2}}\int_{0}^{m(s,s^{\prime})} d\tau\vert t+\gamma-\tau\vert^{-\frac{1}{2}}
                (\cos(2\sqrt{c_{0}}k\sqrt{t+\gamma-\tau})
                )^{-2} \cr
                =i\hbar u_{t-s}u_{t-s^{\prime}}c_{0}\frac{1}{k}(\tan(2\sqrt{c_{0}}k\sqrt{t-m(s,s^{\prime})+\gamma})
                -\tan(2\sqrt{c_{0}}k\sqrt{t+\gamma}))
               \end{array}\end{equation}   Here
               \begin{equation}
               u_{0}=\frac{1}{\sqrt{\gamma}}\cos(2\sqrt{c_{0}}k\sqrt{\gamma}).\end{equation}
               The limit $\gamma\rightarrow 0$ of $u_{t}$ does not
               exist at $ t=0$. Then, the evolution kernel (27)
               is not defined.

               For eq.(38)
              \begin{equation}\begin{array}{l}
              G_{ss^{\prime}}
               =i\hbar u_{t-s}u_{t-s^{\prime}}c_{0}\frac{1}{k}(\cot(2\sqrt{c_{0}}k\sqrt{t+\gamma})
               -\cot(2\sqrt{c_{0}}k\sqrt{t-m(s,s^{\prime})+\gamma}))\end{array}
                \end{equation}
        The limit $\gamma\rightarrow 0$ of $u_{0}$ in eqs.(38),(44) and in the kernel
        (27)  is $2\sqrt{c_{0}}k$. When $t\rightarrow 0$ and $\gamma\rightarrow 0$
then $m(s,s^{\prime})\rightarrow 0$ and $
G_{ss^{\prime}}\rightarrow 0$ in eq.(34).

                 For a negative time in eq.(39) we obtain
               \begin{equation}\begin{array}{l}
               G_{ss^{\prime}}=\hbar u^{(-)}_{t-s}u^{(-)}_{t-s^{\prime}}  c_{0}^{\frac{3}{2}}
               \int_{0}^{m(s,s^{\prime})} d\tau\vert t+\gamma+\tau\vert^{-\frac{1}{2}}
                 (\cosh(2\sqrt{c_{0}}k\sqrt{-t-\gamma-\tau}) )^{-2}\cr
                 =\hbar u^{(-)}_{t-s}u^{(-)}_{t-s^{\prime}} c_{0}\frac{1}{k}
                 (\tanh(2\sqrt{c_{0}}k\sqrt{-t-m(s,s^{\prime})-\gamma})
                 -\tanh(2\sqrt{c_{0}}k\sqrt{-t-\gamma})).\end{array}
                \end{equation}
                $u_{0}$ has no limit when $\gamma\rightarrow 0$.
                Hence, we have a similar problem as at eq.(43).

                For $u^{(-)}_{t}$ of eq.(40) we have

                \begin{equation}\begin{array}{l}
                G_{ss^{\prime}}
                  =\hbar u^{(-)}_{t-s}u^{(-)}_{t-s^{\prime}} c_{0}\frac{1}{k}(\coth(2\sqrt{c_{0}}k\sqrt{-t-\gamma})
                  \cr-\coth(2\sqrt{c_{0}}k\sqrt{-t-m(s,s^{\prime})-\gamma})).
                 \end{array}\end{equation} The limit $\gamma\rightarrow 0$ of $u_{0}$ in eq.(46)
                 exists. Hence, eq.(46) defines the evolution
                 kernel (27) also in the limit $\gamma\rightarrow
                 0$.
        For $k\rightarrow 0$ the correlation $G_{ss^{\prime}}(k)$ is singular
        at small k
                 but this singularity is cancelled by the volume element $d{\bf k}$
                 in the definition of the evolution kernel (27).

We can now recapitulate the findings of this section. For a
positive time $\Gamma$ is real. Hence, $\psi_{t}^{g}$ is  (
unnormalizable) pure phase. $G_{t}$ (28) (as well as
$G_{ss^{\prime}}$) is purely imaginary. For negative time $a$ is
purely imaginary $a\simeq i\sqrt{\vert t+\gamma\vert}$. Then,
\begin{equation}
i\Gamma_{t}=c_{0}^{-\frac{3}{2}}\vert
t+\gamma\vert^{\frac{3}{2}}(u_{t}^{(-)})^{-1}\partial_{t}u_{t}^{(-)}
\end{equation}\
is a real function. We have calculated (47) for the solutions
(37)-(41). It comes out that $i\Gamma$ is negative (hence
$\psi_{t}^{g}$ is normalizable) only for the solution (40) when we
get
\begin{equation}\begin{array}{l}
ic_{0}^{\frac{3}{2}}\vert
t+\gamma\vert^{-\frac{3}{2}}\Gamma_{t}=\frac{1}{2}\vert
t+\gamma\vert^{-1} -\vert
t+\gamma\vert^{-\frac{1}{2}}\sqrt{c_{0}}k\coth(2\sqrt{c_{0}}k\sqrt{\vert
t+\gamma\vert})<0
\end{array}\end{equation}for $\vert t+\gamma\vert >0$. For small  $\vert t+\gamma\vert $
we have $i\Gamma_{t}\simeq -c_{0}^{-\frac{1}{2}}k^{2} \vert
t+\gamma\vert^{\frac{3}{2}}   $  . Hence, $\psi_{t}^{g}$ for the
solution (40) becomes normalizable for negative time. We may say
that there is a smooth limit $\gamma\rightarrow 0$ of quantum
scalar field theory while the metric passes from positive to
negative signature if $\Gamma$ in the WKB state (13) is determined
by the classical wave function solution (38) (for $t+\gamma>0$) or
(40) (for $t+\gamma<0$). $G_{t}$ in the evolution kernel (27) is
purely imaginary for a positive time ,$G_{0}=0$ and $G_{t}$
becomes real and positively definite for a negative time. The
dynamics is well-defined for the negative time in all cases
(39)-(41). However, if $u_{t}$ has no limit $\gamma\rightarrow 0$
and $t\rightarrow 0$ then $u_{0}$ in the evolution kernel (27) is
infinite in the limit of degenerate metric. In these cases a
definition of the evolution of the scalar quantum field during a
change of the signature would need further investigation ( a
possible renormalization removing the infinite constant). The
solution (38) (for positive time) and its continuation to (40)
(for negative time) leads to a continuation of the Gaussian WKB
solution for positive time to a Gaussian normalizable wave
function for the negative time.

\section{Summary}
There is a longstanding question about the relation between the
singularities in Einstein gravity and their role in quantum
theory. The singularity can be expressed in terms of particle's
geodesics in an external metric. The geodesics (under some
positive energy conditions) cannot be extended far to the past.
The relativistic quantum  description of a particle without spin
is treated by quantum scalar field theory. This is a quantization
of the  wave equation for the scalar field $\Phi$ in an external
metric. We define QFT of a scalar field as a solution of the
functional Schr\"odinger equation (determined by the canonical
Hamiltonian) for the wave function $\psi_{t}(\Phi)$. The question
arises as whether we can define a solution of the Schr\"odinger
equation for $t$ far in the past (we can choose the time
coordinate so that the past is defined by $t<0$). We discuss the
scalar field in a background metric of classical radiation. We
consider the initial condition for the Schr\"odinger equation in
the WKB form. Then, (as expected) quantum field theory is
expressed by classical data concerning the wave equation in an
external metric.  The solution $a^{2}=t$ ($t\in R$)
 of the Friedmann equations for classical radiation leads to
a degenerate metric at $t=0$ and is acausal for $t<0$ (being
Euclidean). Such a metric is admissible in a semi-classical
approach to quantum gravity when the solution of the Schr\"odinger
equation is obtained as an average over all saddle points of the
classical action in eq.(1). We discuss the consequences of
extending the Schr\"odinger equation to $t\leq 0$ for the initial
wave function which is of the WKB form (13)-(14). We show that the
WKB phase $\Gamma_{t}$ (13) is defined by a solution of the
classical wave equation in an external metric, whereas the
evolution of the wave function $\psi_{t}(\Phi)$ is determined by
the randomly perturbed wave equation (19)  ( a differentiation of
eq.(19) leads to the wave equation as discussed in sec.4). The
solution of the stochastic equation defines a random field
$\Phi_{t}$ whose correlation functions are related to the
correlation functions of the quantum field by eq.(20). We obtain
explicit solutions of $\Phi_{t}$ and calculate its correlation
functions $G_{t}$ (28). Using this correlation function we can
obtain the evolution kernel (26) for the Schr\"odinger time
evolution. In this sense the quantum time evolution is defined by
classical data: the scalar wave equation and its Green functions.
We can follow the dependence of the classical data on the
background metric. We formulate the initial value problem at $t=0$
for the Schr\"odinger equation whereas we shift the time evolution
of the radiation background as $a^{2}=t+\gamma$ ($\gamma>0$ for
$t>0$ , $\gamma<0$ for $t<0$) in order to separate the classical
singularity of the Friedmann solution and the quantum initial
value problem. We are interested in the behavior of the solution
$\psi_{t}(\Phi)$ of the Schr\"odinger equation close to the
classical singularity $\gamma\rightarrow 0$ . We obtain exact
expressions for the solutions of the scalar wave equation in the
radiation background and for the scalar field correlations $G_{t}$
for all $t\in R$. These results allow us to gain conclusions on
the behavior of the solution of the quantum Schr\"odinger equation
close to the classical singularity at $\gamma=0$. There is a
solution $u_{t}$ of the scalar wave equation in the background
metric $a^{2}=t+\gamma$ ((38) for $t+\gamma>0 $ and (40) for
$t+\gamma<0$ ) which together with its derivative has the limit
$\gamma\rightarrow 0$. This solution determines the WKB wave
function (13) and the stochastic field $\Phi_{t}$ which lead to
the solution of the Schr\"odinger equation $\psi_{t}(\Phi)$ with a
smooth limit at the singularity $\gamma\rightarrow 0$. The
solution of the wave equation in the external metric $a^{2}=t$ for
$t\geq 0$ propagates like in the Minkowski space time in a causal
and oscillatory way. The behavior of the classical solution
$u_{t}$ of the wave equation has its impact on the form of the WKB
solution of the Schr\"odinger equation. When $\gamma=0$ for $t>0$
then the WKB factor is a Gaussian pure phase ($\Gamma_{t}$ in
eq.(13) is real), $\Gamma_{0}=0$, while $i\Gamma_{t}<0$
 for $t<0$ (eqs.(47)-(48)). For $t<0$ the
 solution (1) of the Schr\"odinger equation is a normalizable wave function which behaves
 like a solution
 of a diffusion equation resembling the quantum decay  of the
 wave function after a barrier penetration. The result is an
 outcome of the imaginary value of $\sqrt{-g}$ in eq.(1) and in the Hamiltonian (6)(for
 $t<0$). The time evolution can also be expressed by an evolution kernel (29).
 The evolution kernel is determined by the classical solution
 $u_{t}$ of the wave equation and the correlation function $G_{t}$ of the
 stochastic field $\Phi_{t}$. For a positive time $G_{t}$ is
 purely imaginary. In such a case the evolution kernel is a pure phase in
 agreement with the naive Feynman formula (1).
For a negative time $G_{t}$ is a real function. In such a case
eq.(29) gives an expression for a transition function of a
 diffusion process.
We can conclude that in spite of the degenerate metric we can find
solutions $u_{t}$ of the wave equation varying continuously for
all $t\in R$. As a consequence the  WKB solution $\psi_{t}$ of the
Schr\"odinger equation  for $t\geq 0$ can be continued to a
solution of a diffusion equation  for  $t\leq 0$. The damping of
the classical waves at $t<0$ has its effect on the solution of the
Schr\"odinger equation for $t<0$ which describes a classical
diffusion for a negative time.

  \end{document}